\documentclass[submission, Phys]{SciPost}

\usepackage{braket}
\usepackage{bbm}
\usepackage{bm}
\usepackage{dsfont}
\usepackage{graphicx,graphics}
\usepackage{soul}
\usepackage[makeroom]{cancel}
\usepackage{amsmath,amssymb}
\usepackage{array}
\usepackage{changepage}
\usepackage{url}
\usepackage{hyperref}

\DeclareMathOperator*{\argmin}{argmin}
\DeclareMathOperator*{\argmax}{argmax}

\usepackage[normalem]{ulem} 
\usepackage{xcolor} 
\usepackage{listings}

\definecolor{codegreen}{rgb}{0,0.6,0}
\definecolor{codegray}{rgb}{0.5,0.5,0.5}
\definecolor{codepurple}{rgb}{0.58,0,0.82}
\definecolor{backcolour}{rgb}{1,1,1}

\lstdefinestyle{mystyle}{
    backgroundcolor=\color{backcolour},
    commentstyle=\color{codegreen},
    keywordstyle=\color{magenta},
    numberstyle=\tiny\color{codegray},
    stringstyle=\color{codepurple},
    basicstyle=\ttfamily\footnotesize,
    breakatwhitespace=false,
    breaklines=true,
    captionpos=b,
    keepspaces=true,
    numbers=left,
    numbersep=5pt,
    showspaces=false,
    showstringspaces=false,
    showtabs=false,
    tabsize=2
}

\usepackage{amsmath}

\begin{document}

\begin{center}{\Large \textbf{QGOpt: Riemannian optimization for quantum technologies}}\end{center}

\begin{center}
I. A. Luchnikov\textsuperscript{1,2,3*},
A. Ryzhov\textsuperscript{2},
S. N. Fillipov\textsuperscript{1,4,5}
H. Ouerdane\textsuperscript{2}
\end{center}

\begin{center}
{\bf 1} Moscow Institute of Physics and Technology, Institutskii Pereulok 9, Dolgoprudny, Moscow Region 141700, Russia
\\
{\bf 2} Center for Energy Science and Technology, Skolkovo Institute of Science and Technology, Moscow 121205, Russia
\\
{\bf 3} Russian Quantum Center, Skolkovo, Moscow 143025, Russia
\\
{\bf 4} Steklov Mathematical Institute of Russian Academy of Sciences, Gubkina Street 8, Moscow 119991, Russia
\\
{\bf 5} Valiev Institute of Physics and Technology of Russian Academy of Sciences, Nakhimovskii Prospect 34, Moscow 117218, Russia
\\
* Ilia.Luchnikov@skoltech.ru
\end{center}

\begin{center}
\today
\end{center}


\section*{Abstract}
{\bf
Many theoretical problems in quantum technology can be formulated and addressed as constrained optimization problems. The most common quantum mechanical constraints such as, e.g., orthogonality of isometric and unitary matrices, CPTP property of quantum channels, and conditions on density matrices, can be seen as quotient or embedded Riemannian manifolds. This allows to use Riemannian optimization techniques for solving quantum-mechanical constrained optimization problems. In the present work, we introduce QGOpt, the library for constrained optimization in quantum technology. QGOpt relies on the underlying Riemannian structure of quantum-mechanical constraints and permits application of standard gradient based optimization methods while preserving quantum mechanical constraints. Moreover, QGOpt is written on top of TensorFlow, which enables automatic differentiation to calculate necessary gradients for optimization. We show two application examples: quantum gate decomposition and quantum tomography.
}

\vspace{10pt}
\noindent\rule{\textwidth}{1pt}
\tableofcontents\thispagestyle{fancy}
\noindent\rule{\textwidth}{1pt}
\vspace{10pt}

\textit{Note added}. After the paper was published, Alexander
Pechen brought relevant references to our attention in October
2021: in Ref.~\cite{pechen-2008} the authors proposed an idea to
use complex Stiefel manifolds for parameterizing quantum channels
and performing the gradient optimization for quantum control and
quantum technologies; in Ref.~\cite{oza-2009} the authors
developed an approach to optimization of quantum systems with an
arbitrary finite dimension for quantum control and quantum
technologies via gradient flows over complex Stiefel manifolds.

\section{Introduction}
\label{sec:intro}
Many quantum-mechanical problems can be solved using optimization methods as illustrated by the following examples. The ground state of a quantum system with Hamiltonian $H$ can be found using the variational method, which is akin to an optimization problem \cite{toulouse2016introduction}:
\begin{equation}
    \ket{\Omega} = \argmin_{\ket{\psi}}\frac{\bra{\psi} H \ket{\psi}}{\braket{\psi|\psi}},
\end{equation}
where  $\ket{\psi}$ is a non-normalized trial state, $\ket{\Omega}$ is the non-normalized ground state. This formulation of a ground state search problem was successfully used for the study of many-body quantum systems \cite{ceperley1977monte, bressanini2002robust}. In particular, the ground state of a correlated spin system can be found in the following forms: matrix product states \cite{schollwock2011density, perez2006matrix, orus2014practical}, projected entangled pair states \cite{verstraete2008matrix, verstraete2004renormalization} or neural networks \cite{carleo2017solving, hibat2020recurrent, choo2019two}. To perform variational energy optimization one can utilize optimization algorithms such as the density matrix renormalization group \cite{white1992density, schollwock2005density}, the time evolving block decimation \cite{vidal2003efficient, vidal2004efficient, orus2008infinite} for tensor network architectures, the quantum natural gradient \cite{stokes2020quantum}, and adaptive first-order optimization methods like the Adam optimizer \cite{kingma2014adam} for neural-networks-based quantum parametrization.

Problems of reconstruction of quantum states, quantum channels and quantum processes from measured data can also be formulated as optimization problems. For example, the state of a many-body quantum system can be reconstructed with neural networks by maximization of the logarithmic likelihood function on a set of measurement outcomes \cite{torlai2018neural, cha2020attention, carrasquilla2019reconstructing, luchnikov2019variational}. The Choi matrix of an unknown quantum channel can be reconstructed in a tensor network form via the minimization of the Kullback-Leibler divergence \cite{torlai2020quantum}. Non-Markovian quantum dynamics can be reconstructed from measured data in different ways \cite{luchnikov2020machine, banchi2018modelling} by use of optimization algorithms.

Some quantum mechanics problems require keeping certain constraints while minimizing or maximizing an objective function. For example, quantum phase transitions can be described using an entanglement renormalization technique, which requires an optimization over matrices with orthogonality constraints, i.e. isometric matrices. To solve this problem, Vidal and Evenbly suggested an algorithm \cite{evenbly2009algorithms, evenbly2014algorithms, vidal2007entanglement} that does not have analogs in standard optimization theory. Another example of a constrained optimization problem emerging in quantum mechanics is quantum channel tomography. It requires preservation of natural ``quantum'' constraints, i.e. the completely positive and trace preserving (CPTP) property of quantum channels \cite{holevo2012quantum}. Constraints preservation here can be achieved by using a particular parametrization or by adding regularizers that ensure that the constraints are satisfied. 

Adding regularizers into a loss function merely provides approximate preservation of constraints, and a naive parametrization may lead to over-parametrization and result in the optimization slowing down. One therefore needs a universal approach to quantum technology optimization. As many natural ``quantum'' constraints can be seen as Riemannian manifolds, Riemannian optimization can become a candidate well-suited for the role of universal framework for constrained optimization in quantum mechanics. In the present work, we introduce QGOpt (Quantum Geometric Optimization) \cite{QGOpt_repo}, a library for Riemannian optimization in quantum mechanics and quantum technologies. It allows one to perform an optimization with all typical constraints of quantum mechanics.

This article is organized as follows. In Sec. 2, we give an overview of Riemannian optimization. We then turn to Riemannian manifolds in quantum mechanics in Sec. 3. In Sec. 4, we present the QGOpt application programming interface (API), and we illustrate its use in Sec. 5, with two examples: quantum gate decomposition and quantum channel tomography. In Sec. 6, we also show that QGOpt can handle optimization over an arbitrary Cartesian product of manifolds.

\section{Overview of the Riemannian optimization}
\label{sec:another}
While optimizing an objective function defined on the Euclidean space, one performs a sequence of elementary operations like points and vectors transportation. We call these elementary operations optimization primitives. For example, one iteration of the simplest gradient descent method involves an update of the current estimation of the optimal point $x_t$ as follows: $x_{t+1} = x_{t} + v_t$, where $v_t = -\eta \nabla f(x_t)$ is a vector tending to improve the current estimation, $t$ is the number of previous iterations, and $\eta$ is the step-size. This update can be seen as a transportation of a point $x_t$ along a vector $v_t$. More sophisticated algorithms may require keeping additional information about the optimization landscape in terms of vectors $\{m_t^{(0)},\dots,m_t^{(N)}\}$ associated with the current point $x_t$. These vectors should be transported together with $x_t$ to a new point and then updated according to a particular algorithm. However, as transportation of vectors in a Euclidean space is the trivial identity transformation, it may be safely skipped. Optimization on curved spaces requires a generalization of optimization primitives in a certain way. As an example of optimization algorithms we consider a gradient descent with momentum \cite{ruder2016overview} and its Riemannian generalization \cite{becigneul2018riemannian, li2020efficient}. We keep our overview simple, and for an in-depth introduction to the topic, we recommend references \cite{boumal2020introduction, absil2009optimization}.

Let us assume that we aim to minimize the value of a function $f:\mathbb{R}^n\to \mathbb{R}$, and that we have access to its gradient $\nabla f(x)$. In the Euclidean space $\mathbb{R}^n$, a gradient descent with momentum consists of the following steps wrapped into a loop:
\begin{enumerate}
    \item Calculation of the momentum vector $m_{t+1} = \beta m_{t} + (1 - \beta)\nabla f(x_t)$,
    \item Taking a step along the direction of a momentum vector $x_{t+1} = x_t - \eta m_{t+1}$,
\end{enumerate}
where the initial momentum vector $m_0$ is the null vector, $\beta$ is a hyperparameter whose value is usually taken around $\beta \approx 0.9$, and $\eta$ is the size of the optimization step.

Let us assume now that a function $f$ is defined on a Riemannian manifold ${\cal M}$ that is embedded in the Euclidean space: $f:{\cal M}\to \mathbb{R}$. Then we can no longer apply the standard scheme of gradient descent with momentum because it clearly takes $x_t$ out of the manifold ${\cal M}$. This scheme can be generalized step by step. First, we have to generalize the notion of a gradient. The standard Euclidean gradient is not a tangent vector to a manifold and it does not take into account the metric of a manifold. One may introduce the Riemannian gradient that can be constructed based on the standard gradient $\nabla f(x)$. The Riemannian gradient lies in the space tangent to a point $x$ and properly takes the metric of a tangent space into account. Although an optimization algorithm takes a step along a vector tangent to a manifold, it still takes a point out of the manifold. In order to fix this issue, one can replace a straight line step with a proper curved line step. In the Riemannian geometry the generalization of the straight line step is given by the notion of exponential map that reads
\begin{equation}
    x_{\rm out} = {\rm Exp}_{x_{\rm in}}(v) = \gamma (1),
\end{equation}
where $\gamma(t)$ is a geodesic \cite{pokorny2012geodesics} such that $\gamma(0) = x_{\rm in}$ and $\frac{d\gamma(t)}{dt}\big|_{t=0} = v$, $x_{\rm in}$ is an initial point on a manifold, $x_{\rm out}$ is a final point. However, in practice the calculation of a geodesic is often computationally too inefficient. In these cases, one can use a retraction instead of an exponential map, which is a first-order approximation of a geodesic \cite{pokorny2012geodesics}:
\begin{equation}
    \tilde{x}_{\rm out} = R_{x_{\rm in}}(v),
\end{equation}
where $\tilde{x}_{\rm out}$ also lies in a manifold and $\|\tilde{x}_{\rm out} - x_{\rm out}\|=O(\|v\|^2)$. A retraction is not unique so it can be chosen to be computationally efficient.

The gradient descent with momentum also requires to transport the momentum vector at each iteration from a previous point to a new point. The Euclidean version of the gradient descent with momentum does not have an explicit step with transportation of the momentum vector because in the Euclidean space transportation of a vector is trivial. However, this step is necessary in the Riemannian case, where the trivial Euclidean vector transportation takes a vector out of a tangent space. A vector transport $\tau_{x, w}(v)$ is the result of transportation of a vector $v$ along a vector $w$ which takes into account that a tangent space varies from one manifold's point to another in the Riemannian case. The overall Riemannian generalization of the gradient descent with momentum can be summarized as follows:
\begin{enumerate}
    \item Calculation of the momentum vector $\tilde{m}_{t+1} = \beta m_{t} + (1 - \beta)\nabla_R f(x_t)$,
    \item Taking a step along a new direction of the momentum $x_{t+1} = R_{x_t}(- \eta \tilde{m}_{t+1})$,
    \item Transport of the momentum vector to a new point $x_{t+1}$: $m_{t+1} = \tau_{x_t, -\eta \tilde{m}_{t+1}}(\tilde{m}_{t+1})$.
\end{enumerate}
Note that other first-order optimization methods can be generalized in a similar fashion.

\section{Riemannian manifolds in quantum mechanics}
\label{section:manifolds}
Many objects of quantum mechanics can be seen as elements of smooth manifolds. However, their mathematical description, suitable for numerical algorithms, may involve some abstract constructions that should be clarified. In this section we consider an illustrative example of the set of Choi matrices and describe this set as a smooth quotient manifold. We restrict our consideration to a plain description of all necessary mathematical concepts. At the end of the section, we also list all the manifolds implemented in the QGOpt library and describe their possible use.

The evolution of any quantum system that interacts with its environment can be described by a quantum channel.
Here, we consider quantum channels defined as the following CPTP linear map:  $\Phi:\mathbb{C}^{n\times n}\to \mathbb{C}^{n\times n}$. Any quantum channel can be represented through its Choi matrix \cite{holevo2012quantum}. A Choi matrix is a positive semi-definite operator $C\in\mathbb{C}^{n^2\times n^2}$ that has a constraint ${\rm Tr}_p(C) = {\mathds 1}$, where ${\rm Tr}_p$ is a partial trace over the first subsystem and ${\mathds 1}$ is the identity matrix. To make the notion of the partial trace less abstract, let us consider a piece of the TensorFlow code, which computes a partial trace of a Choi matrix. First, we apply a reshape operation to a Choi matrix that changes the shape of a matrix as follows
\begin{lstlisting}[language=Python, numbers=none]
C_resh = tf.reshape(C, (n, n, n, n))
\end{lstlisting}
The tensor $C_{\rm resh}\in \mathbb{C}^{n\times n\times n\times n}$ is an alternative representation of the Choi matrix. Further in the text, we distinguish two equivalent representations of a Choi matrix: $C$ and $C_{\rm resh}$. The partial trace of a Choi matrix can be calculated using $C_{\rm resh}$ as follows $[{\rm Tr}_p(C)]_{i_1i_2} = \sum_{j}[C_{\rm resh}]_{i_1ji_2j}$. Practically it can be done by running
\begin{lstlisting}[language=Python, numbers=none]
tf.einsum('ikjk->ij', C_resh)
\end{lstlisting}
which means that we take a trace over the first and third indices (with numeration of indices starting from 0).

\begin{figure}
    \centering
    \includegraphics[scale=0.47]{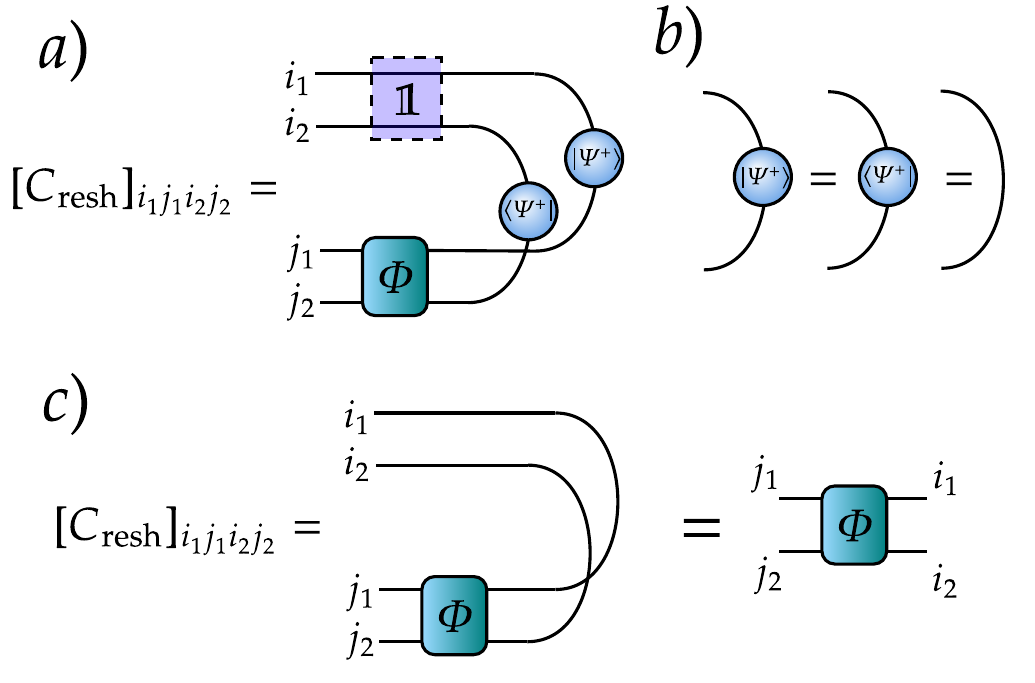}
    \caption{a) Diagrammatic representation of the Choi matrix. The block denoted by $\mathds{1}$ represents the identity map in the definition of the Choi matrix. b) One can note that the state of a two-component quantum system $\ket{\Psi^+}$ can be seen as the identity matrix. c) Finally, we note that the Choi matrix is a quantum channel itself.}
    \label{fig:choi_iso}
\end{figure}

The Choi–Jamio{\l}kowski isomorphism \cite{jamiolkowski1972linear} establishes a one-to-one correspondence between quantum channels and Choi matrices. One can calculate the Choi matrix of a known quantum channel as follows
\begin{equation}
    C = ( \mathds{1}\otimes \Phi ) \ket{\Psi^+}\bra{\Psi^+},
    \label{eq:choi_matrix}
\end{equation}
where $\ket{\Psi^+} = \sum_{i=1}^n \ket{i}\otimes \ket{i}$ and $\{\ket{i}\}_{i=1}^n$ is an orthonormal basis in $\mathbb{C}^{n}$. In order to show that the Choi matrix is a quantum channel itself, we consider the representation of Eq.~(\ref{eq:choi_matrix}) in terms of tensor diagrams \cite{bridgeman2017hand,Biamonte2017}. The reshaped version of a Choi matrix $[C_{\rm resh}]_{i_1j_1i_2j_2}$ is shown in Fig.~\ref{fig:choi_iso}. The tensor diagrams in Fig.~\ref{fig:choi_iso} show that $|\Psi^+\rangle$ and $\mathds{1}$ in the definition of the Choi matrix lead only to relabeling of multi-indices.

\begin{figure}
    \centering
    \includegraphics[scale=0.47]{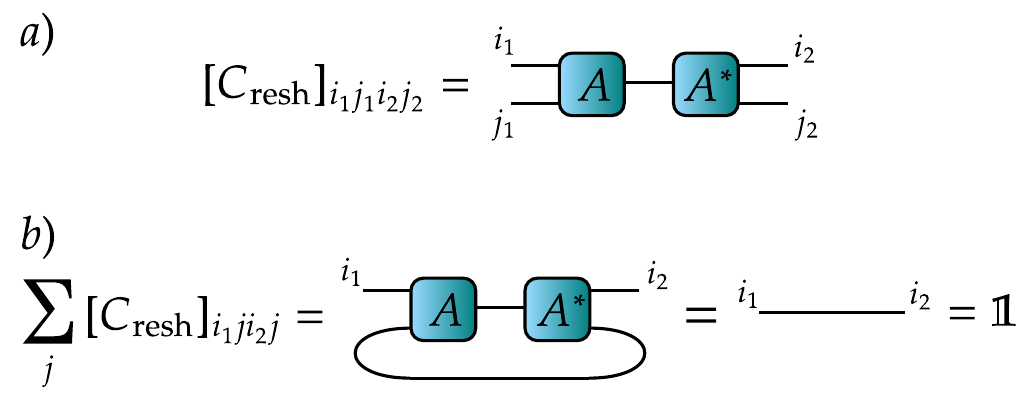}
    \caption{a) Decomposition of a Choi matrix into $A$ and $A^\dagger$. b) Diagrammatic representation of the isometric property of $A$.}
    \label{fig:choi_decomposition}
\end{figure}

\begin{figure}
    \centering
    \includegraphics[scale=0.47]{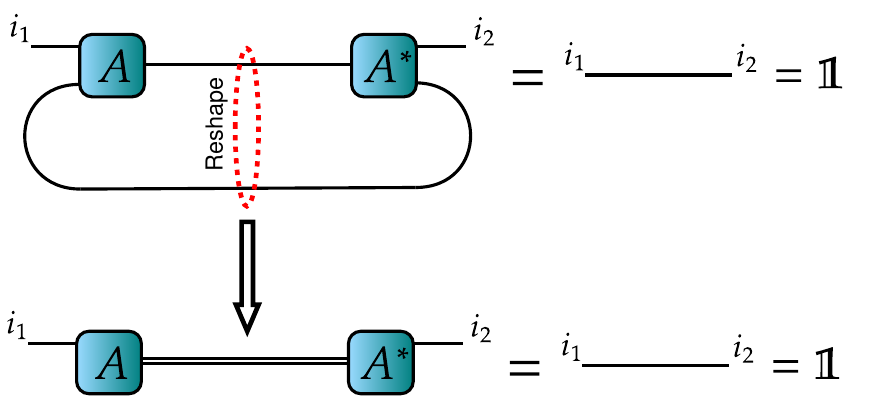}
    \caption{Diagrammatic representation of the reshape operation turning the tensor $A_{\rm resh}$ into an isometric matrix.}
    \label{fig:reshaped_A}
\end{figure}

The set of all Choi matrices of size $n^2 \times n^2$ (the corresponding quantum channel acts on density matrices of size $ n\times n$) $C_n$ is the following subset of $\mathbb{C}^{n^2\times n^2}$
\begin{equation}
   C_n=\left\{C\in \mathbb{C}^{n^2\times n^2}\big|C\geq 0, \ {\rm Tr}_p(C)=\mathds{1}\right\},
\end{equation}
where $C\geq 0$, and ${\rm Tr}_p(C)=\mathds{1}$ corresponds to the CPTP property of the corresponding quantum channel.
This subset can be described as a Riemannian manifold that admits different Riemannian optimization algorithms. In order to describe $C_n$ as a Riemannian manifold, we may parametrize the Choi matrix with an auxiliary matrix $A \in \mathbb{C}^{n^2\times n^2}$:
\begin{equation}
    \label{eq:AC_connection}
    C = AA^\dagger.
\end{equation}
The matrix $C$ is positive semi-definite by construction. We also distinguish $A\in \mathbb{C}^{n^2\times n^2}$ and its reshaped version $A_{\rm resh}\in \mathbb{C}^{n \times n \times n^2}$ that are connected by the reshape operation. The condition on a partial trace of a Choi matrix transforms to the following equality:
\begin{equation}
\label{eq:isometric_a}
    [{\rm Tr}_p(C)]_{i_1i_2} = [{\rm Tr}_p(A^\dagger A)]_{i_1i_2} = \sum_{kj} [A_{\rm resh}]^*_{ki_1j}[A_{\rm resh}]_{ki_2j} = \delta_{i_1i_2},
\end{equation}
and its diagrammatic form is given in Fig.~\ref{fig:choi_decomposition}.
One can see that if in Eq.~\eqref{eq:isometric_a} we recast the two indices $k$ and $j$ into one index $q$, we then end up with the following relation:
\begin{equation}
    \sum_q [A_{\rm resh}]^*_{qi_1}[A_{\rm resh}]_{qi_2} = \delta_{i_1i_2},
    \label{eq:reshaped_A}
\end{equation}
which means that $[A_{\rm resh}]_{qi}$ is an isometric matrix and the corresponding tensor $[A_{\rm resh}]_{kij}$ is a reshaped isometric matrix. The corresponding diagrammatic representation of Eq.~\eqref{eq:reshaped_A} is given in Fig.~\ref{fig:reshaped_A}. We call such a tensor obtained by reshaping an isometric matrix an isometric tensor.
The set of all complex isometric matrices of fixed size forms a Riemannian manifold called complex Stiefel manifold \cite{edelman1998geometry} that we denote as ${\rm St}$. Equations~\eqref{eq:isometric_a} and~\eqref{eq:reshaped_A}, and the diagram Fig.~\ref{fig:reshaped_A} show that the set of tensors $A_{\rm resh}$ can be seen as a complex Stiefel manifold.

\begin{figure}
    \centering
    \includegraphics[scale=0.47]{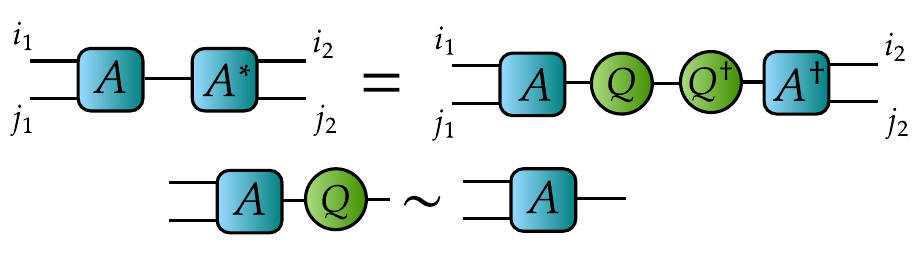}
    \caption{Diagrammatic representation of Eq.~\eqref{eq:gauge_invariance}.}
    \label{fig:gauge_invariance}
\end{figure}

\begin{figure}
    \centering
    \includegraphics[scale=0.45]{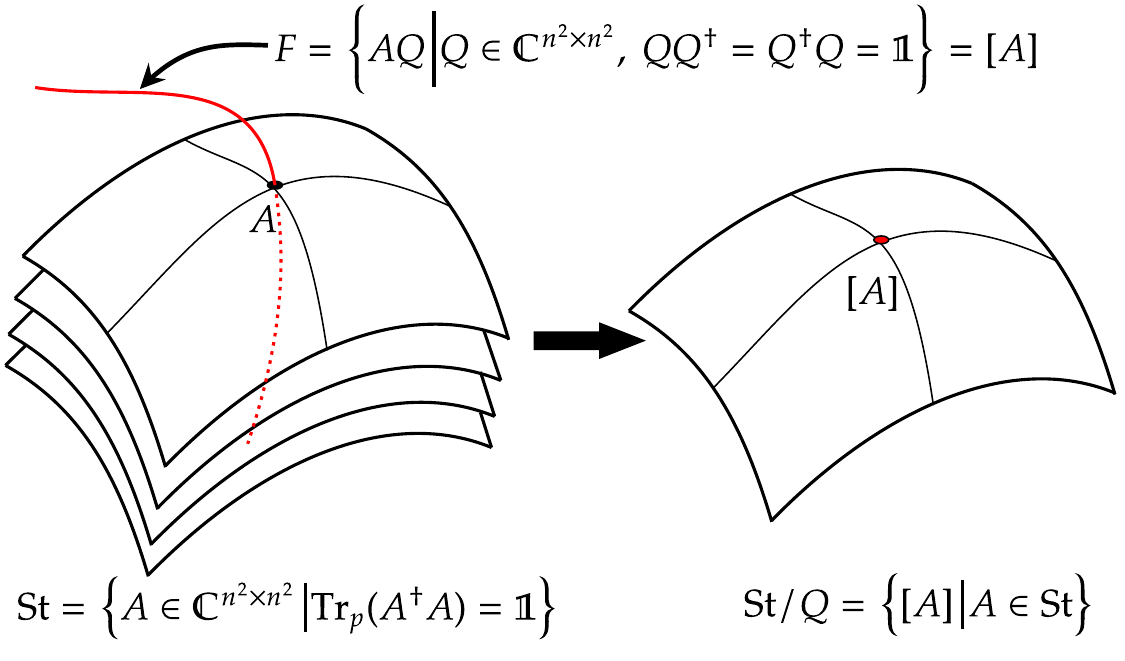}
    \caption{Graphical representation of the transition from the manifold ${\rm St}$ of all matrices $A$, to the quotient manifold ${\rm St/Q}$ that eliminates undesirable symmetry of the parametrization. The red curve represents a particular equivalence class $F$ that is also called a fiber.}
    \label{fig:quotient_structure}
\end{figure}

At first glance, it looks like we have shown that the set of Choi matrices can be seen as a Stiefel manifold, but there is a problem that invalidates this statement: the matrices $A$ and $AQ$, where $Q$ is an arbitrary unitary matrix, correspond to the same Choi matrix; in other words, we have an equivalence relation $AQ \sim A$. Indeed,
\begin{equation}
    C = A Q Q^\dagger A^\dagger = A A^\dagger.
    \label{eq:gauge_invariance}
\end{equation}
A diagrammatic version of Eq.~\eqref{eq:gauge_invariance} is depicted in Fig.~\ref{fig:gauge_invariance}.
It shows that for any $A$ there is a family of equivalent matrices $[A] = \{AQ|Q\in\mathbb{C}^{n^2\times n^2}, \ Q^\dagger Q = QQ^\dagger = \mathds{1}\}$, which is the equivalence class of $A$ and leads to the same Choi matrix. One can eliminate this symmetry by turning to a quotient manifold ${\rm St} / Q = \left\{[A]|A\in{\rm St}\right\}$, which consists of equivalence classes. This rather abstract construction can be imagined as a projection of a manifold along surfaces representing equivalence classes (see Fig.~\ref{fig:quotient_structure}). Having a map $\pi(A) = [A]$ and a horizontal lift \cite{boumal2020introduction}, that connects tangent spaces of ${\rm St}/Q$ and tangent spaces of ${\rm St}$, one can describe the abstract manifold ${\rm St}/Q$ through ${\rm St}$. The quotient manifold ${\rm St}/Q$ can be further identified with the set of Choi matrices $C_n$. It allows one to perform a Riemannian optimization on $C_n$, by using the parametrization $C = AA^\dagger$. Mathematical details of this construction are given in Appendix \ref{app:mathematical_details}.

The example of the quotient manifold representing the Choi matrices through their parametrization shows all the necessary steps that emerge while building the mathematical description of quantum mechanical manifolds. The set of all manifolds implemented in QGOpt library is listed below.
\begin{itemize}
    \item The complex Stiefel manifold $\displaystyle {\rm St}_{n,p} =\left\{V\in \mathbb{C}^{n\times p} |V^{\dagger } V=\mathds{1}\right\}$ is a set of all isometric matrices of fixed size. A particular case of this manifold is a set of all unitary matrices of fixed size; therefore, this manifold can be used for different tasks related to quantum control. Some architectures of tensor networks may include isometric matrices as building blocks \cite{luchnikov2020riemannian, hauru2020riemannian}; thus, one can use this manifold to optimize such tensor networks.

    \item The manifold of density matrices of fixed rank\\ $\varrho_{n, r}=\left\{\varrho \in \mathbb{C}^{n\times n}\Bigl| \varrho =\varrho ^{\dagger } ,\ {\rm Tr}(\varrho )=1,\ \varrho \geq 0, \ {\rm rank}(\varrho)=r\right\}$ is a set of all fixed-rank Hermitian positive semi-definite matrices with unit trace. Since density matrices represent states of quantum systems, one can use this manifold to perform state tomography and optimization of initial quantum states in different quantum circuits. This manifold is implemented through a parametrization with a quotient structure on top of it.

    \item The manifold of Choi matrices of fixed rank\\ $C_{n, r} =\left\{C\in \mathbb{C}^{n^{2} \times n^{2}}\Bigl| C=C^{\dagger } ,\ {\rm Tr}_{p} (C)=\mathds{1},\ C\geq 0, \ {\rm rank(C) = r}\right\}$ is a set of all fixed-rank Hermitian positive semi-definite matrices with auxiliary linear constraint (equality of the partial trace to the identity matrix). Choi matrices are used as representations of quantum channels; hence, one may use this manifold to perform quantum channel tomography and optimization of quantum channels in different quantum circuits. This manifold is implemented through a parametrization with a quotient structure on top of it.

    \item The manifold of Hermitian matrices $\displaystyle H_{n} =\left\{H\in \mathbb{C}^{n\times n}\Bigl| H=H^{\dagger }\right\}$ is a linear subspace of a space $\mathbb{C}^{n\times n}$. Since Hermitian matrices represent measurable physical operators in the quantum theory, one can use this manifold to perform a search of optimal measurable physical operators in different problems.

    \item The manifold of Hermitian positive definite matrices $\displaystyle \mathbb{S}^{n}_{++} =\left\{S\in \mathbb{C}^{n\times n}\Bigl| S=S^{\dagger } ,\ S\succ 0\right\}$ is a set of all positive definite matrices of fixed size. One can use it to search the optimal non-normalized quantum state in different tasks.

    \item The manifold of positive operator-valued measures (POVMs) with full rank elements {\footnotesize $\displaystyle {\rm POVM}_{m,n} =\left\{\{E_{i} \}^{m}_{i=1} \in \mathbb{C}^{m\times n\times n}\Biggl| E_{i} =E^{\dagger }_{i} ,\ E_{i} \geq 0,\ \sum ^{m}_{i=1} E_{i} =\mathds{1}, \ {\rm rank}(E_i)=n\right\}$} can be considered as a tensor with Hermitian positive semi-definite full-rank slices that sum into the identity matrix. Since POVMs describe generalized measurements in quantum theory, one can use this manifold to perform a search of optimal measurements that give the largest information gain. This manifold is implemented through a parametrization with a quotient structure on top of it.
\end{itemize}
Mathematical details of the implementation of manifolds are given in Appendix \ref{app:mathematical_details}.

\section{QGOpt API}
\subsection{Manifolds API}
In this section we discuss the API of the version 1.0.0 of the QGOpt library.
The central class of the QGOpt library is the manifold base class. All particular manifold types are inherited from the manifold base class. All manifold subclasses admit working with the direct product of several manifolds. Optimization primitives of each particular manifold are implemented as methods of the corresponding class describing a manifold. This list of methods allows one not to pay particular attention to the details of the underlying Riemannian geometry.

Let us consider basic illustrative examples. First, one needs to import all necessary libraries and create an example of a manifold. As an example we consider the complex Stiefel manifold.
\begin{lstlisting}[language=Python]
import QGOpt as qgo
import tensorflow as tf

# example of complex Stiefel manifold
m = qgo.manifolds.StiefelManifold()
\end{lstlisting}
Here, $m$ is an example of the complex Stiefel manifold that contains all the necessary information on the manifold's geometry. Some manifolds allow one to specify a type of metric and retraction as well. Using this example of a manifold one can sample a random point from a manifold:
\begin{lstlisting}[language=Python, numbers=none]
u = m.random((4, 3, 2))
\end{lstlisting}
Here, we sample a random tensor $u$, that is a complex valued TensorFlow tensor of size $4\times 3\times 2$. This tensor represents a point from the direct product of four complex Stiefel manifolds. The first index of this tensor enumerates a manifold and the last two indices are matrix indices. Therefore, the tensor $u$ can be seen as a set of four isometric matrices. One can generate a random tangent vector drawn from $u$.
\begin{lstlisting}[language=Python, numbers=none]
v = m.random_tangent(u)
\end{lstlisting}
Here, $v$ is a complex-valued TensorFlow tensor of the same size and type as $u$, and represents the random tangent vector drawn from $u$. Now let us assume that we have a random vector $w$ which is of the same size and type, but is not tangent to $u$. One can make the orthogonal projection of this vector on the tangent space of $u$:
\begin{lstlisting}[language=Python, numbers=none]
w = m.proj(u, w)
\end{lstlisting}
The updated vector $w$ is an element of the tangent space of $u$ now. The projection method of quotient manifold performs the projection on the horizontal space. To get the scalar product of two tangent vectors one can use the following line of code:
\begin{lstlisting}[language=Python, numbers=none]
wv_inner = m.inner(u, w, v)
\end{lstlisting}
Here we pass $u$ to the inner product method to specify the tangent space where we compute the inner product, because in Riemannian geometry the metric and inner product are point-dependent in general.

To implement first-order Riemannian optimization methods on a manifold one needs to be able to move points and vectors along the manifold. There are retraction and vector transport methods for this purpose. As an example let us move a point $u$ along a tangent vector $v$ via the retraction map:
\begin{lstlisting}[language=Python, numbers=none]
u_tilde = m.retraction(u, v)
\end{lstlisting}
The new point $\tilde{u}$ is the result of transportation of $u$ along vector $v$.
To perform transportation of a vector along some other vector one can run:
\begin{lstlisting}[language=Python, numbers=none]
v_tilde = m.vector_transport(u, v, w)
\end{lstlisting}
Here we start from point $u$ and transport a tangent vector $v$ along a tangent vector $w$, and obtain $\tilde{v}$ that is the result of the vector transportation.

The last important method converts the Euclidean gradient of a function to the Riemannian gradient. The Riemannian gradient replaces the Euclidean gradient to take into account the metric of a manifold and the tangent space in a given point. To calculate the Riemannian gradient one can use:
\begin{lstlisting}[language=Python, numbers=none]
r = m.egrad_to_rgrad(u, e)
\end{lstlisting}
where we denote the Euclidean gradient as $e$ and the Riemannian gradient as $r$.

The numerical complexity of each optimization primitive varies from one manifold to another. The complexity of all primitives is summarized in Appendix \ref{app:complexity}.

\subsection{Optimizers}
The Riemannian optimizers implemented in QGOpt are inherited from TensorFlow optimizers and hence have the same API. The main difference is that one should also pass an example of manifold while defining an optimizer, which guides the optimizer and preserves the manifold's constraints. Two optimizers are implemented, that are among the most popular in machine learning: Riemannian versions of Adam \cite{kingma2014adam} and SGD \cite{10.2307/2236626}.

If $m$ is a manifold element and lr is a learning rate (optimization step size), then the Adam and SGD optimizers can be initialized as:

\begin{lstlisting}[language=Python]
# Riemannian ADAM optimizer
opt = qgo.optimizers.RAdam(m, lr)
# Riemannian SGD optimizer
opt = qgo.optimizers.RSGD(m, lr)
\end{lstlisting}

\noindent Note that some other attributes like the momentum value of the SGD optimizer or the AMSGrad modification of the Adam optimizer can also be specified.

\subsection{Auxiliary functions}
It is important to keep in mind that TensorFlow optimizers work well only with real variables. Therefore, one cannot use complex variables to represent a point on a manifold because they are being tuned while optimizing. The simplest way of representing a point from a complex manifold through real tensors is by introducing an additional index that enumerates real and imaginary parts of a  tensor. For  example a complex-valued tensor of shape $(a,b,c)$ can be represented as a real-valued tensor of shape $(a,b,c,2)$. During calculations, we need to convert tensors from their real representation to their complex representation and back.

Let us initialize a complex-valued tensor as a point from a manifold using the method ``random''. In order to make this tensor a  variable suitable for an optimizer, one needs to convert it to the real representation. Then, while building a computational graph, one may need to have a complex form of a tensor again:

\begin{lstlisting}[language=Python]
# a random real tensor, last index enumerates
# real and imaginary parts
w = tf.random.normal((4, 3, 2),
                     dtype=tf.float64)
# corresponding complex tensor of shape (4, 3)
wc = qgo.manifolds.real_to_complex(w)
# corresponding real tensor (wr = w)
wr = qgo.manifolds.complex_to_real(wc)
\end{lstlisting}

\section{Examples of application of QGOpt}
\subsection{Quantum gate decomposition}
In this subsection we consider an illustrative example of quantum gate decomposition. It is known that any two qubit-quantum gate $U$ can be decomposed \cite{shende2004minimal}:
\begin{equation}
    U = [\tilde{u}_{11}\otimes \tilde{u}_{12}] U_{\rm CNOT} [\tilde{u}_{21}\otimes \tilde{u}_{22}] \times U_{\rm CNOT} [\tilde{u}_{31}\otimes \tilde{u}_{32}] U_{\rm CNOT} [\tilde{u}_{41}\otimes \tilde{u}_{42}],
\end{equation}
where $U_{\rm CNOT}$ is the CNOT gate and $\{\tilde{u}_{ij}\}_{i,j=1}^{4, 2}$ is a set of unknown one qubit-gates. Since a set $\{\tilde{u}_{ij}\}_{i,j=1}^{4, 2}$ can be seen as the direct product of $8$ complex Stiefel manifolds, one can use Riemannian optimization methods to find all $\tilde{u}_{ij}$. First, we initialize randomly a trial set $\{u_{ij}\}_{i,j=1}^{4, 2}$ that will be tuned by Riemannian optimization methods. For simplicity, we denote the decomposition introduced above as
\begin{equation}
   D\left(u_{ij}\right) = [u_{11}\otimes u_{12}] U_{\rm CNOT} [u_{21}\otimes u_{22}] \times U_{\rm CNOT} [u_{31}\otimes u_{32}] U_{\rm CNOT} [u_{41}\otimes u_{42}].
\end{equation}
The optimal set of one-qubit gates can be expressed as:
\begin{eqnarray}
    \argmin_{\{u_{ij}\}_{i,j=1}^{4, 2}} \|U - D(u_{ij})\|_F,
\end{eqnarray}
where each $u_{ij}$ obeys the unitarity constraint and $\| \cdot \|_F$ is the Frobenius norm.

\begin{figure}
    \centering
    \includegraphics[scale=0.6]{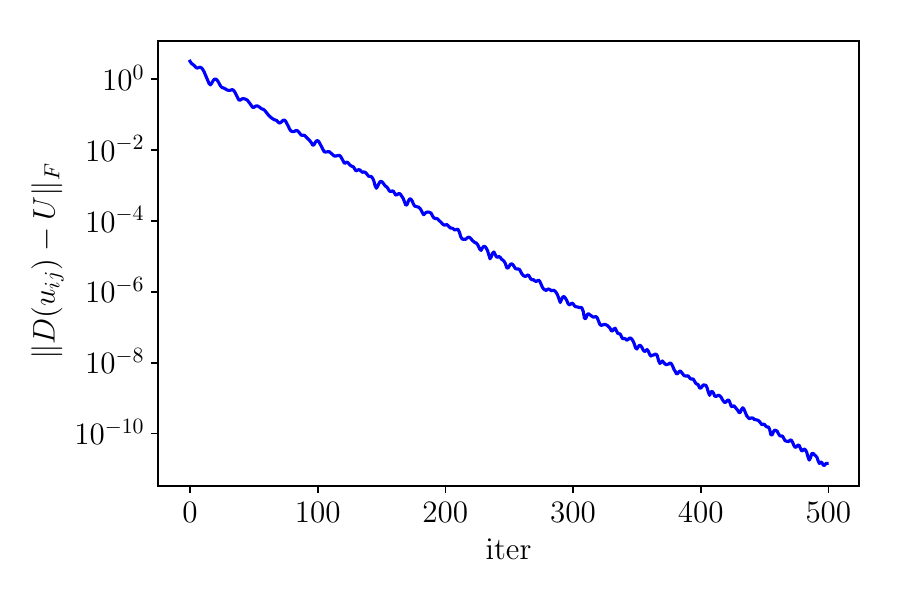}
    \caption{Frobenius distance between a gate and its decomposition. One can see that the distance rapidly decreases with the number of iteration towards nearly zero within machine precision.}
    \label{fig:gate_decomposition}
\end{figure}

Before considering the main part of the code that solves the above problem, we need to introduce a function that calculates the Kronecker product of two matrices:
\begin{lstlisting}[language=Python]
def kron(A, B):
    AB = tf.tensordot(A, B, axes=0)
    AB = tf.transpose(AB, (0, 2, 1, 3))
    AB = tf.reshape(AB, (A.shape[0]*B.shape[0],
                         A.shape[1]*B.shape[1]))
    return AB
\end{lstlisting}
Now, we define an example of the complex Stiefel manifold:
\begin{lstlisting}[language=Python, numbers=none]
m = qgo.manifolds.StiefelManifold()
\end{lstlisting}
We use a randomly generated target gate that we want to decompose,
\begin{lstlisting}[language=Python, numbers=none]
U = m.random((4, 4), dtype=tf.complex128)
\end{lstlisting}
We initialize the initial set $\{u_{ij}\}_{i,j=1}^{4, 2}$ randomly as a 4th rank tensor,
\begin{lstlisting}[language=Python, numbers=none]
u = m.random((4, 2, 2, 2), dtype=tf.complex128)
\end{lstlisting}
The first two indices of this tensor enumerate a particular one-qubit gate, the last two indices are matrix indices of a gate. We turn this tensor into its real representation in order to make it suitable for an optimizer and wrap it up into the TensorFlow variable:
\begin{lstlisting}[language=Python, numbers=none]
u = qgo.manifolds.complex_to_real(u)
u = tf.Variable(u)
\end{lstlisting}
We initialize the CNOT gate $U_{\rm CNOT}$ as follows:
\begin{lstlisting}[language=Python]
cnot = tf.constant([[1, 0, 0, 0],
                    [0, 1, 0, 0],
                    [0, 0, 0, 1],
                    [0, 0, 1, 0]],
                    dtype=tf.complex128)
\end{lstlisting}
As the next step, we initialize the Riemannian Adam optimizer:
\begin{lstlisting}[language=Python, numbers=none]
lr = 0.2  # optimization step size
opt = qgo.optimizers.RAdam(m, lr)
\end{lstlisting}
and run the forward pass of computations:
\begin{lstlisting}[language=Python]
with tf.GradientTape() as tape:
    # turning u back into its
    # complex representation
    uc = qgo.manifolds.real_to_complex(u)
    # decomposition
    D = kron(uc[0, 0], uc[0, 1])
    D = cnot @ D
    D = kron(uc[1, 0], uc[1, 1]) @ D
    D = cnot @ D
    D = kron(uc[2, 0], uc[2, 1]) @ D
    D = cnot @ D
    D = kron(uc[3, 0], uc[3, 1]) @ D
    # loss function
    L = tf.linalg.norm(D - U) ** 2
    # is equivalent to casting to a real dtype
    L = tf.math.real(L)
\end{lstlisting}
The final step is to minimize the loss function $L = \|D(u_{ij}) - U\|_F^2$ calculated during the previous step. We calculate the gradient of $L$, using automatic differentiation, with respect to the set $\{u_{ij}\}_{i,j=1}^{4, 2}$:
\begin{lstlisting}[language=Python, numbers=none]
grad = tape.gradient(L, u)
\end{lstlisting}
and pass the gradient to the optimizer:
\begin{lstlisting}[language=Python, numbers=none]
opt.apply_gradients(zip([grad], [u]))
\end{lstlisting}
The Adam optimizer performs one optimization step keeping the orthogonality constraints. We repeat the forward pass, gradient calculation and optimization steps several times, wrapping them into a for loop until convergence and end up with a proper decomposition of the gate $U$. The optimization result is given in Fig.~\ref{fig:gate_decomposition}. One can see that at the end of the optimization process, the error is completely negligible. This section in the form of a tutorial is available in the QGOpt online \cite{QGOpt_docs}.

\subsection{Quantum tomography}
Another typical problem that can be addressed by Riemannian optimization is the quantum tomography of states \cite{blume2010optimal, teo2016introduction} and channels \cite{knee2018quantum, mohseni2008quantum}. Here, we consider an example of quantum tomography of channels because it involves a more complicated structure than quantum tomography of states.

Let ${\cal H} = \bigotimes_{i=1}^n \mathbb{C}^2$ be the Hilbert space of a system consisting of $n$ qubits.
Let us assume that one has a set of input states $\{\rho_i\}_{i=1}^N$, where $N$ is a total number of states, and each $\rho_i$ is a density matrix on ${\cal H}$. One passes initial states through an unknown quantum channel $\Phi_{\rm true}$ and observes a set of measurement outcomes $\left\{M^{\rm tetra}_{k_i^1}\otimes \dots\otimes M^{\rm tetra}_{k_i^n}\right\}_{i=1}^N$, where $M^{\rm tetra}_k$ is an element of a tetrahedral POVM \cite{renes2004symmetric}:
\begin{eqnarray}
&&M_{k}^{\rm tetra} = \frac{1}{4}\left({\mathds 1} + \bm{s}_k^T\bm{\sigma}\right),\ k \in (0, 1, 2, 3), \\
&& \bm{\sigma} = \left(\sigma_x, \sigma_y, \sigma_z\right),\nonumber s_0 = (0, 0, 1),s_1 = \left(\frac{2\sqrt{2}}{3}, 0, -\frac{1}{3}\right), \\ &&s_2 = \left(-\frac{\sqrt{2}}{3}, \sqrt{\frac{2}{3}}, -\frac{1}{3}\right), s_3 = \left(-\frac{\sqrt{2}}{3}, -\sqrt{\frac{2}{3}}, -\frac{1}{3}\right).\nonumber
\end{eqnarray}
One can estimate an unknown channel by maximizing the logarithmic likelihood of measurement outcomes:
\begin{eqnarray}
\argmax_{\Phi \text{\ is \ CPTP}} \sum_{i=1}^N\log\left(M^{\rm tetra}_{k_i^1}\otimes \dots\otimes M^{\rm tetra}_{k_i^n}\Phi(\rho_i)\right).
\end{eqnarray}
For simplicity, we assume that the many-body tetrahedral POVM $M$ is already predefined and has the shape $(2^{2n}, 2^n, 2^n)$, where the first index enumerates the POVM element. We also assume that we have a data set that consists of a set of initial density matrices of shape $(N, 2^n, 2^n)$ and a set of POVM elements of the same shape that came true after measurements. In our experiments, the unknown channel has Kraus rank $2$ and is generated randomly, the initial density matrices are pure and also generated randomly.

Let us proceed with the practical implementation. First, we define an example of the quotient manifold equivalent to the manifold of Choi matrices:
\begin{lstlisting}[language=Python, numbers=none]
m = qgo.manifolds.ChoiMatrix()
\end{lstlisting}

\noindent Elements of this manifold are connected with Choi matrices via the relation~\eqref{eq:AC_connection}. We randomly initialize a point from the quotient manifold,\\

\begin{figure}
    \centering
    \includegraphics[scale=0.6]{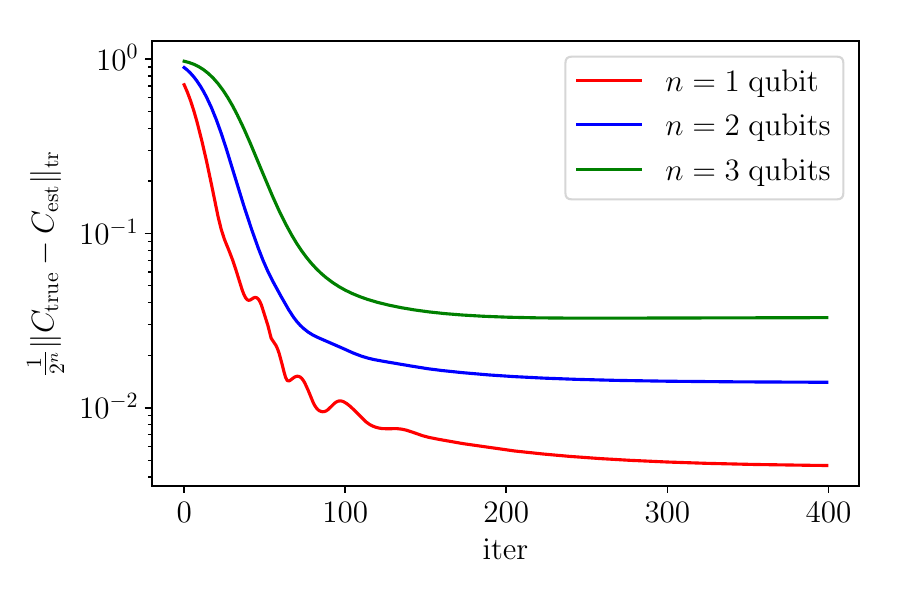}
    \caption{Dependence between Jamio{\l}kowski process distance and number of iteration. Number of measurement outcomes $N=600000$ for all experiments.}
    \label{fig:channel_tomography}
\end{figure}

\begin{lstlisting}[language=python]
# random initial parametrization
A = m.random((2**(2*n), 2**(2*n)),
dtype=tf.complex128)
# variable should be real
# to make an optimizer work correctly
A = qgo.manifolds.complex_to_real(A)
# variable
A = tf.Variable(A)
\end{lstlisting}
Then we initialize the Riemannian Adam optimizer:
\begin{lstlisting}[language=Python, numbers=none]
lr = 0.07
opt = qgo.optimizers.RAdam(m, lr)
\end{lstlisting}
and calculate the logarithmic likelihood function:
\begin{lstlisting}[language=python]
with tf.GradientTape() as tape:
    # Ac is a complex representation of A
    # shape=(2**2n, 2**2n)
    Ac = qgo.manifolds.real_to_complex(A)

    # reshape parametrization
    # (2**2n, 2**2n) --> (2**n, 2**n, 2**2n)
    Ac = tf.reshape(Ac, (2**n, 2**n, 2**(2*n)))

    # Choi tensor (reshaped Choi matrix)
    choi = tf.tensordot(Ac,
                        tf.math.conj(Ac),
                        [[2], [2]])

    # turning Choi tensor to the
    # corresponding quantum channel
    phi = tf.transpose(choi, (1, 3, 0, 2))
    phi = tf.reshape(phi, (2**(2*n), 2**(2*n)))

    # reshape initial density
    # matrices to vectors
    rho_resh = tf.reshape(rho_in, (N, 2**(2*n)))

    # passing density matrices
    # through a quantum channel
    rho_out = tf.tensordot(phi,
                           rho_resh,
                           [[1], [1]])
    rho_out = tf.transpose(rho_out)
    rho_out = tf.reshape(rho_out,
                         (N, 2**n, 2**n))

    # probabilities of measurement outcomes
    # (povms is a set of POVM elements
    # came true of shape (N, 2**n, 2**n))
    p = tf.linalg.trace(povms @ rho_out)

    # negative log likelihood (to be minimized)
    L = -tf.reduce_mean(tf.math.log(p))
\end{lstlisting}
The complexity of the code above can be reduced by choosing the optimal order of tensor contraction; however, it becomes more complicated in this case, and is not suitable for the tutorial.
Finally, we calculate the logarithmic likelihood gradient with respect to the parametrization of the Choi matrix:
\begin{lstlisting}[language=Python, numbers=none]
grad = tape.gradient(L, A)
\end{lstlisting}
and apply the optimizer to make an optimization step that does not violate the CPTP constraints:
\begin{lstlisting}[language=Python, numbers=none]
opt.apply_gradients(zip([grad], [A]))
\end{lstlisting}
We repeat the calculation of the logarithmic likelihood function, gradient calculation and optimization steps several times, wrapping them into a for loop, until convergence is reached. To evaluate the quality of an unknown quantum channel estimation, we calculate the Jamio{\l}kowski process distance \cite{gilchrist2005distance}:
\begin{equation}
J(\Phi_{\rm true}, \Phi_{\rm est}) = \frac{1}{2^n}\|C_{\rm true} - C_{\rm est}\|_{\rm tr},
\end{equation}
where $\Phi_{\rm true}(\Phi_{\rm est})$ is the true (estimated) quantum channel, $C_{\rm true}(C_{\rm est})$ is the corresponding Choi matrix, $\|\cdot\|_{\rm tr}$ is the trace norm and $0 \leq J(\Phi_{\rm true}, \Phi_{\rm est}) \leq 1$. One can see in Fig.~\ref{fig:channel_tomography} that the Jamio{\l}kowski process distance converges to some small value with the number of iterations and we end up with a reasonable estimation of an unknown quantum channel. This section is available in the QGOpt online documentation in the form of a tutorial \cite{QGOpt_docs}.

\section{Optimization over an arbitrary Cartesian product of manifolds.}
In general, it is possible to perform optimization over the Cartesian product of different manifolds. The QGOpt library allows finding
\begin{equation}
    \argmin_{A\in{\cal M}} f(A),
\end{equation}
where ${\cal M}$ is an arbitrary Cartesian product of manifolds, implemented in the QGOpt library, $f$ is a function that can be evaluated within the TensorFlow framework.

As an example, let us consider an optimization over the manifold
\begin{equation}
    {\cal M} = \varrho_{n_1, n_1} \times \varrho_{n_2 \times n_2}\times \varrho_{n_2 \times n_2}\times C_{n, r}\times C_{n, r},
\end{equation}
where $\times$ denotes the Cartesian product. In other words, one has one manifold of full-rank density matrices of size $n_1 \times n_1$, two manifolds of full-rank density matrices of size $n_2 \times n_2$ and two manifolds of Choi matrices of size $n^2 \times n^2$ and rank $r$. Let us define the building blocks of ${\cal M}$,
\begin{lstlisting}[language=python, numbers=none]
    m_choi = qgo.manifolds.ChoiMatrix()
    m_dens = qgo.manifolds.DensityMatrix()
\end{lstlisting}
The next step is to define variables representing points on manifolds. First, we define a variable representing a point in $\varrho_{n_1, n_1}$
\begin{lstlisting}[language=python]
# random initialization
A_rho_1 = m_dens.random((n1, n1),
                        dtype=tf.complex128)
# variable should be real
# to make an optimizer work correctly
A_rho_1 = qgo.manifolds.complex_to_real(A_rho_1)
# variable
A_rho_1 = tf.Variable(A_rho_1)
\end{lstlisting}
Then we define a variable representing a point from $\varrho_{n_2, n_2}\times \varrho_{n_2, n_2}$
\begin{lstlisting}[language=python]
# random initialization
A_rho_2 = m_dens.random((2, n2, n2),
                        dtype=tf.complex128)
# variable should be real
# to make an optimizer work correctly
A_rho_2 = qgo.manifolds.complex_to_real(A_rho_2)
# variable
A_rho_2 = tf.Variable(A_rho_2)
\end{lstlisting}
where we take advantage of the fact that both matrices are of the same size, and we can represent them as one tensor. Let us group these two variables into one list
\begin{lstlisting}[language=python, numbers=none]
A_rho = [A_rho_1, A_rho_2]
\end{lstlisting}
which is passed to the Riemannian optimizer on the manifold of density matrices. One also needs to define a variable representing a point from $C_{n, r}$.
\begin{lstlisting}[language=python]
# random initialization
A_choi = m_choi.random((n**2, r))
# variable should be real
# to make an optimizer work correctly
A_choi = qgo.manifolds.complex_to_real(A_choi)
# variable
A_choi = tf.Variable(A_choi)
\end{lstlisting}
Now one needs to define optimizers
\begin{lstlisting}[language=python]
# learning rate
lr = 0.01
# optimizer over density matrices
opt_dens = qgo.optimizers.RAdam(m_dens, lr)
# optimizer over choi matrices
opt_choi = qgo.optimizers.RAdam(m_choi, lr)
\end{lstlisting}
and perform an optimization step
\begin{lstlisting}[language=python]
with tf.GradientTape() as tape:
    L = f(A_rho_1, A_rho_2, A_choi)
# gradient over all variables
grad_total = tape.gradient(L, A_rho + [A_choi])
# gradient over variables representing
# density matrices
grad_rho = grad_total[:2]
# gradient over the variable representing
# Choi matrix
grad_choi = grad_total[-1]
# optimization step
opt_dens.apply_gradients(zip(grad_rho, A_rho))
opt_choi.apply_gradients(zip([grad_choi],
                             [A_choi]))
\end{lstlisting}
where we assume that the function $f$ is predefined. In order to iterate optimization steps until convergence, one can wrap the code above into a loop. This general scheme can be used for optimization over an arbitrary set of different manifolds.

\section{Discussion and concluding remarks}
The range of application of the QGOpt library to different problems of quantum technologies is not limited to quantum gate decomposition and quantum tomography. The six manifolds implemented in QGOpt give rise to different interesting scenarios of constrained optimization usage in quantum technologies. For example, the complex Stiefel manifold can be used to address different control problems \cite{glaser2015training, goerz2019krotov, goerz2017charting} where one needs to find an optimal set of unitary gates driving a quantum system to a desirable quantum state. It is also possible to use a complex Stiefel manifold to perform entanglement renormalization \cite{luchnikov2020riemannian, hauru2020riemannian}, machine learning by unitary tensor networks \cite{liu2019machine} or non-Markovian quantum dynamics identification \cite{luchnikov2020machine}. In addition, quotient manifolds for quantum tomography, or for density matrices and Choi matrices can be used to maintain natural quantum constraints in different tensor network architectures. Quotient manifold of POVMs can be used for searching for an optimal generalized measurement scheme with maximum information gain. Finally, all these manifolds can be combined in one optimization task, which allows to address multi-component problems.

Although as of yet the QGOpt library includes only first-order optimization methods, we plan to extend the list of optimizers by including quasi-Newton methods such as the Riemannian BFGS \cite{huang2016riemannian}, and the recently developed quantum natural gradient descent \cite{stokes2020quantum} generalized to the case of embedded and quotient manifolds.

To conclude, we have introduced the QGOpt library aimed at solving constrained optimization problems with natural quantum constraints. We have introduced and discussed abstract concepts such as quotient manifolds under the hood of QGOpt. We have gone through the QGOpt API and covered its most important features. We also sorted out examples of codes solving illustrative quantum technology problems.

\section*{Acknowledgements}
The authors thank Stephen Vintskevich and Mikhail Krechetov for fruitful discussions. The authors also thank the anonymous Referees and Dr. Michael H. Goerz who provided very useful reports on the manuscript.


\paragraph{Funding information}
 I.A.L. and S.N.F. thank the Foundation for the Advancement of Theoretical Physics and Mathematics ``BASIS'' for support under Project No. 19-1-2-66-1.

\begin{appendix}

\section{Underlying geometry of manifolds implemented in the QGOpt library}
\label{app:mathematical_details}

In this appendix, we consider some mathematical aspects of the implementation of manifolds in the QGOpt library. First, we discuss how one can identify complex matrices, which are elements of all manifolds implemented in the QGOpt library, with real matrices. Any complex matrix $A$ can be represented as follows
\begin{equation}
    \tilde{A} = \begin{bmatrix} {\rm Re}(A) & {\rm Im}(A) \\ -{\rm Im}(A) & {\rm Re}(A) \end{bmatrix}.
\end{equation}
The following correspondences between operations with complex matrices and operations with their real representations
\begin{equation}
    A^\dagger \rightarrow \tilde{A}^T, \ AB \rightarrow \tilde{A}\tilde{B}, \ A + B \rightarrow \tilde{A} + \tilde{B}, \ 2{\rm Re}({\rm Tr}(A)) = {\rm Tr}(\tilde{A}),
\end{equation}
allow us to work with certain sets of complex matrices as with Riemannian manifolds of real matrices \cite{sato2014riemannian}. The QGOpt library contains six manifolds: three implemented as embedded manifolds and three implemented as quotient manifolds.

Table \ref{table:embedded_manifolds} summarizes the geometry lying under the hood of a high-level description of the embedded manifolds in the QGOpt library.

\begin{table}
    \centering
    \scriptsize
    \rotatebox{90}{
    \begin{tabular}{| m{2cm} | m{4cm} | m{4cm} | m{4cm} | m{3cm} | m{3cm} |}
     \hline
     Manifold & Description & Inner product & Riemannian gradient & Retraction & Vector transport \\
     \hline
     Complex Stiefel manifold ${\rm St}_{n, p}$ &
     Embedded manifold of complex isometric matrices. & Two types of inner product \cite{edelman1998geometry, luchnikov2020riemannian}, induced by embedding, are available in the QGOpt library: Eucledian inner product $\langle v, w \rangle_u = {\rm Re}({\rm Tr}(v^\dagger w))$, and canonical inner product $\langle v, w \rangle_u = {\rm Re}\left({\rm Tr}\left(v^\dagger \left(I - \frac{1}{2}uu^\dagger\right)w\right)\right)$. These two types of inner product induce the same orthogonal projection \cite{edelman1998geometry}. & The Riemannian gradient for the Euclidean inner product takes the following form \cite{edelman1998geometry} $\nabla_R f(u) = \frac{1}{2}u(u^\dagger \nabla f(u) - \nabla f(u)^\dagger u) + (I - uu^\dagger)\nabla f(u)$, the Riemannian gradient for the canonical inner product takes the following form \cite{edelman1998geometry} $\nabla_R f(u) = \nabla f(u) - u\nabla f(u)^\dagger u$. & Three types of retraction are available in the QGOpt library: SVD decomposition based retraction \cite{absil2009optimization}, QR decomposition based retraction \cite{absil2009optimization} and Cayley retraction \cite{absil2009optimization, li2020efficient}. & Vector transport is induced by a retraction. It is implemented as the orthogonal projection of a vector on the tangent space of a point obtained via a retraction. \cite{absil2009optimization, li2020efficient}.\\
     \hline
     Manifold of Hermitian positive definite matrices $\mathbb{S}_{++}^n$ & Embedded manifold of complex Hermitian, positive definite matrices. Two  different inner products are introduced in a way that the manifold is complete. Inner products are not extended on the ambient space, which does not allow the orthogonal projection on the tangent space of a point. However, for this manifold we do not use the orthogonal projection at all. & Two types of inner product are available in the QGOpt library: Log--Cholesky inner product \cite{lin2019riemannian, luchnikov2020riemannian} and Log--Euclidean inner product \cite{luchnikov2020riemannian}. Both inner products keep the manifold complete. & The Riemannian gradient for both inner products that are used in the QGOpt library is derived in \cite{luchnikov2020riemannian}. & Instead of a retraction, one uses the exponential map for both inner products in the QGOpt library. Closed form of the exponential map for the Log--Euclidean inner product can be found in \cite{luchnikov2020riemannian}, for the Log--Cholesky inner product in \cite{lin2019riemannian, luchnikov2020riemannian}. & Instead of a vector transport, one uses the parallel transport for both inner products in the QGOpt library. The closed form of the parallel transport for the Log--Euclidean metric can be found in \cite{luchnikov2020riemannian}, for the Log--Cholesky metric in \cite{lin2019riemannian, luchnikov2020riemannian}.\\
     \hline
     Manifold of Hermitian matrices $H_n$. & Embedded manifold of Hermitian matrices that also is a linear subspace of the ambient space. & Only the Euclidean inner product is available in the QGOpt library; it reads $\langle v, w\rangle_u = {\rm Re}({\rm Tr}(v^\dagger w))$. The inner product is induced by the embedding. & The Riemannian gradient for the Euclidean inner product takes the following form $\nabla_R f(u) = \frac{1}{2}\left(\nabla f(u) + \nabla f(u)^\dagger\right)$. & Instead of a retraction, one uses the exponential map, which is a trivial transportation along a straight line. & Instead of a vector transport, one uses the parallel transport, which is the identity transformation.\\
     \hline
    \end{tabular}
    }
    \caption{Summary of the geometry of embedded manifolds implemented in the QGOpt library.}
    \label{table:embedded_manifolds}
\end{table}

~\\
We also summarize the geometry of the manifolds that are implemented as quotient manifolds. In our summary, we follow the book by Nicolas Boumal \cite{boumal2020introduction}, which provides a very instructive presentation of the optimization on quotient manifolds.

Having an optimization problem on a quotient manifold, one works with two sets $\overline{{\cal M}}$ and ${\cal M}$ that are connected as
\begin{equation}
    {\cal M} = {\cal M}/\sim = \{[x]|x\in\overline{\cal M}\},
\end{equation}
where $[x] = \{y|y \in \overline{\cal M}, \ y \sim x\}$ is the equivalence class of $x$, $\overline{\cal M}$ is some Riemannian manifold and ${\cal M}$ is its quotient.
We call a map $\pi$ a canonical projection if it maps any $x$ from $\overline{\cal M}$ to its equivalence class:
\begin{equation}
    \pi(x) = [x].
\end{equation}
For ${\cal M}$ to be a manifold, one requires $\pi$ to be smooth and its differential $D\pi(x):T_{x}\overline{\cal M}\rightarrow T_{[x]}{\cal M}$ must have a constant rank $r = \dim\left({\cal M}\right)$ for all $x \in \overline{\cal M}$. We call $V_x$ a vertical space at $x\in\overline{\cal M}$ if it is the kernel of $D\pi(x)$, i.e.
\begin{equation}
    V_x = {\rm ker}\left(D\pi(x)\right),
\end{equation}
then one can decompose the tangent space $T_x\overline{\cal M}$ at a point $x$ as
\begin{equation}
    T_x\overline{\cal M} = H_x \oplus V_x,
\end{equation}
where $H_x$ is the orthogonal complement of $V_x$ also called the horizontal space. The restricted linear map $D\pi(x)|_{H_x}:H_x\rightarrow T_{[x]}{\cal M}$ is bijective by construction and can be used to represent a vector from $T_{[x]}{\cal M}$ as a vector from $H_x$. This representation is called horizontal lift and reads
\begin{equation}
    v = (D\pi(x)|_{H_x})^{-1}[\xi] = {\rm lift}_x(\xi),
\end{equation}
where $\xi$ is a vector from $T_{[x]}{\cal M}$ and $v$ is its representation from $H_x$.

Having introduced all the objects above, one can try to construct all primitives for optimization algorithms on a quotient manifold through the same primitives on a total manifold (see table \ref{table:primitives}).

\begin{table}[h]
    \centering
    \begin{tabular}{| m{3.5cm} | m{10.5cm} |}
     \hline
        ~& ~\\
     Inner product & $\langle\xi, \zeta \rangle_{[x]} = \langle {\rm lift}_x(\xi), {\rm lift}_x(\zeta) \rangle_x$, where $\langle\cdot,\cdot\rangle_x$ is an inner product in $T_x\overline{\cal M}$ and $\xi,\zeta \in T_{[x]}{\cal M}$\\
     ~& ~\\
     Retraction & $R_{[x]}(\xi) = \pi\left(\overline{R}_x({\rm lift}_x(\xi))\right)$, where $\overline{R}$ is an retraction on $\overline{\cal M}$\\
     ~& ~\\
     Vector transport & $\tau_{[x], \xi}(\zeta) = {\rm lift}^{-1}_{\overline{R}_x({\rm lift}_x(\xi))}\left(P_{H_{\overline{R}_x({\rm lift}_x(\xi))}}({\rm lift}_x(\zeta))\right)$, where $P_S$ is the orthogonal projection operator on a subspace $S$\\
     ~& ~\\
     Function & $\overline{f} = f \circ \pi$, where $f:{\cal M} \rightarrow \mathbb{R}$ and $\overline{f}:\overline{\cal M} \rightarrow \mathbb{R}$\\
     ~& ~\\
     Riemannian gradient & $\nabla_R f([x]) = {\rm lift}^{-1}_x\left(\nabla_R \overline{f}(x)\right)$\\
        ~& ~\\
     \hline
    \end{tabular}
\caption{Optimization primitives of ${\cal M}$ expressed through optimization primitives of $\overline{\cal M}$.}
\label{table:primitives}
\end{table}

These primitives are correct if they do not depend on a choice of a particular point from an equivalence class, i.e. for all $[x]\in{\cal M}$ and $\xi, \ \zeta \in T_{[x]}{\cal M}$ if $x \sim y$ the following statements are true
\begin{equation}
    \label{eq:inner_inv}
    \langle {\rm lift}_x(\xi), {\rm lift}_x(\zeta) \rangle_x = \langle {\rm lift}_y(\xi), {\rm lift}_y(\zeta) \rangle_y,
\end{equation}
\begin{equation}
    \label{eq:retraction_inv}
    \pi\left(\overline{R}_x({\rm lift}_x(\xi))\right) = \pi\left(\overline{R}_y({\rm lift}_y(\xi))\right),
\end{equation}
\begin{equation}
    \label{eq:transport_inv}
    {\rm lift}^{-1}_{\overline{R}_x({\rm lift}_x(\xi))}\left(P_{H_{\overline{R}_x({\rm lift}_x(\xi))}}({\rm lift}_x(\zeta))\right) = {\rm lift}^{-1}_{\overline{R}_y({\rm lift}_y(\xi))}\left(P_{H_{\overline{R}_y({\rm lift}_y(\xi))}}({\rm lift}_y(\zeta))\right),
\end{equation}
where $\sim$ denotes equivalence relation between elements of $\overline{\cal M}$. It is also worth noting that in practice there is no need to go back from $\overline{\cal M}$ to ${\cal M}$ after application of each primitive. Instead, one can work only with objects from $\overline{\cal M}$, which makes optimization algorithms on ${\cal M}$ almost identical to algorithms on $\overline{\cal M}$.

Manifolds $\varrho_{n, r}$, $C_{n, r}$ and ${\rm POVM}_{m, n}$ in the QGOpt library are implemented using the above idea. The quotient geometry of the real version of the manifold $\varrho_{n, r}$ is described in \cite{journee2010low} and also is implemented in the Manopt library \cite{boumal2014manopt}. The alternative approach to optimization on a ${\rm POVM}_{m, n}$ is also considered in \cite{mishra2019riemannian} and implemented in Manopt. To the best of the authors' knowledge, the manifold $C_{n, r}$ has not been considered from the Riemannian optimization point of view.

Let us consider total manifolds that are used to build quotient manifolds implemented in the QGOpt library. They are
\begin{eqnarray}
&&\overline{\varrho}_{n, r} = \left\{A\in \mathbb{C}^{n \times r}_*\big|{\rm Tr}(AA^\dagger)=1\right\},\\
&&\overline{C}_{n, r} = \left\{A\in\mathbb{C}^{n^2\times r}_*\Big|{\rm Tr}_p\left(AA^\dagger\right)=\mathds{1}\right\},\\
&& \overline{{\rm POVM}}_{m, n} = \left\{\{A_i\}_{i=1}^m \Bigg| \sum_{i=1}^m A_iA_i^\dagger = \mathds{1}, \ A_i\in \mathbb{C}^{n \times n}_*\right\},
\end{eqnarray}
where $\mathbb{C}^{p \times q}_*$ is the set of complex full-rank matrices of size $p \times q$.
One can note that the manifold $\overline{\varrho}_{n, r}$ is a sphere with the additional condition on the rank of $A$, manifolds $\overline{C}_{n, r}$ and $\overline{\rm POVM}_{m, n}$ are complex Stiefel manifolds with the additional condition on the ranks of $A$ and $A_i$. Any element of the total manifolds above corresponds to either a density matrix, a Choi matrix, or a POVM. Indeed
\begin{eqnarray}
    &&\varrho = AA^\dagger, \text{ if } A\in \overline{\varrho}_{n, r},\\
    &&C = AA^\dagger, \text{ if } A\in \overline{C}_{n, r},\\
    &&E_i = A_iA_i^\dagger, \text{ if } A\in \overline{\rm POVM}_{m, n},
\end{eqnarray}
where $\varrho$ is some density matrix, $C$ is some Choi matrix and $E_i$ is an element of some POVM. However, there is an ambiguity:
\begin{eqnarray}
    &&\varrho = AA^\dagger = A QQ^\dagger A^\dagger, \\
    &&C = AA^\dagger = A QQ^\dagger A^\dagger, \\
    &&E_i = A_iA^\dagger_i = A_i Q_iQ_i^\dagger A_i^\dagger.
\end{eqnarray}
where $Q$ is unitary and  $\{Q_i\}_{i=1}^m$ is a set of unitary matrices of the appropriate size.
In order to lift the ambiguity we introduce equivalence classes
\begin{eqnarray}
    &&[A] = \{AQ|Q \in \mathbb{C}^{r \times r}, \ QQ^\dagger = \mathds{1}\}\text{, for } \overline{\varrho}_{n, r} \text{ and } \overline{C}_{n, r},\\
    &&[\{A\}_{i=1}^m] = \{\{A_iQ_i\}_{i=1}^m|Q_i \in \mathbb{C}^{n \times n}, \ Q_iQ_i^\dagger = \mathds{1}\}\text{, for } \overline{\rm POVM}_{m, n},
\end{eqnarray}
and the corresponding quotient manifolds
\begin{eqnarray}
    &&\overline{\varrho}_{n, r} / \sim = \left\{[A]\big|A\in \overline{\varrho}_{n, r}\right\},\\
    &&\overline{C}_{n, r} / \sim = \left\{[A]\big|A\in \overline{C}_{n. r}\right\},\\
    &&\overline{\rm POVM}_{m, n} / \sim = \left\{[\{A_i\}_{i=1}^m]\big|\{A_i\}_{i=1}^m \in \overline{\rm POVM}_{m, n}\right\}.
\end{eqnarray}
To identify quotient manifolds with those that are introduced in the main text, we introduce the maps
\begin{eqnarray}
    &&\phi_\varrho:\overline{\varrho}_{n, r}/\sim \rightarrow \varrho_{n, r}:[A] \mapsto AA^\dagger,\\
    &&\phi_C:\overline{C}_{n, r}/\sim \rightarrow C_{n, r}:[A] \mapsto AA^\dagger,\\
    &&\phi_{\rm POVM}:\overline{\rm POVM}_{n, m}/\sim \rightarrow {\rm POVM}_{n, m}:[\{A\}_{i=1}^m] \mapsto \{A_iA^\dagger_i\}_{i=1}^m.
\end{eqnarray}
These three maps are bijections, which follows from Proposition 2.1 in \cite{massart2020quotient} that is proved for real matrices, but generalization to the complex case is straightforward. They, as well as their inverses, are also differentiable which implies that these three maps are diffeomorphisms. It is enough to identify quotient manifolds with those introduced in the main text and turn to the optimization on quotient manifolds.

Now, to perform optimization on $\varrho_{n, r}$, $C_{n, r}$ and ${\rm POVM}_{m, n}$ it is sufficient to introduce appropriate primitives for $\overline{\varrho}_{n, r}$, $\overline{C}_{n, r}$ and $\overline{{\rm POVM}}_{m, n}$ that additionally satisfy Eqs.~(\ref{eq:inner_inv}-\ref{eq:transport_inv}), and the projection on the horizontal space. The total manifolds equipped with the following inner products, induced by inner products of ambient spaces
\begin{eqnarray}
    &&\langle v, w \rangle_A = {\rm Re}\left({\rm Tr}(vw^\dagger)\right), \text{ where } w, \ v \in T_A\overline{\varrho}_{n, r},\\
    &&\langle v, w \rangle_A = {\rm Re}\left({\rm Tr}(vw^\dagger)\right), \text{ where } w, \ v \in T_A\overline{C}_{n, r},\\
    &&\langle \{v_i\}_{i=1}^m, \{w_i\}_{i=1}^m \rangle_A = {\rm Re}\left(\sum_i{\rm Tr}(v_iw_i^\dagger)\right),\\
        \nonumber
        &&\text{ where } \{w_i\}_{i=1}^m, \ \{v_i\}_{i=1}^m \in T_{\{A_i\}_{i=1}^m}\overline{\rm POVM}_{m, n},
\end{eqnarray}
satisfy the condition \eqref{eq:inner_inv} as shown in \cite{boumal2020introduction}. The projections on the horizontal space for the total manifolds are
\begin{eqnarray}
    &&P_{H_A}(v) = P_{T_A \overline{\varrho}_{n, r}}(v) - P_{V_A}(v),\text{ for } \overline{\varrho}_{n, r},\\
    &&P_{H_A}(v) = P_{T_A \overline{C}_{n, r}}(v) - P_{V_A}(v),\text{ for } \overline{C}_{n, r},\\
    &&P_{H_{\{A_i\}_{i=1}^m}}(\{v_i\}_{i=1}^m) = P_{T_{\{A_i\}_{i=1}^m} \overline{\rm POVM}_{n, m}}(\{v_i\}_{i=1}^m) - P_{V_{\{A_i\}_{i=1}^m}}(\{v_i\}_{i=1}^m),\\
        \nonumber
        &&\text{ for } \overline{\rm POVM}_{n, m},
\end{eqnarray}
where the projections on tangent spaces are known for the total manifolds that are the sphere and the complex Stiefel manifolds; and the projections on the vertical spaces can be found by solving the Sylvester equation \cite{Yatawatta:2013xja}. One can introduce several different retractions for total manifolds that, however, may not satisfy the condition Eq.~\eqref{eq:retraction_inv}. Since manifolds $\overline{C}_{n, r}$ and $\overline{\rm POVM}_{m, n}$ are complex Stiefel manifolds we can use SVD-based retraction for them. One can show that SVD-based retraction satisfies the condition Eq.~\eqref{eq:gauge_invariance} (see \cite{boumal2020introduction}). For the manifold $\overline{\varrho}_{n, r}$ one can use retraction on a sphere (see Example 4.1.1 in \cite{absil2009optimization}). This retraction also satisfies the condition Eq.~\eqref{eq:retraction_inv}. Vector transports (see Table  \ref{table:primitives}) induced by retractions above also satisfies the condition Eq.~\eqref{eq:transport_inv}. The Riemannian gradients for $\overline{\varrho}_{n, r}$, $\overline{C}_{n, r}$ and $\overline{\rm POVM}_{m, n}$ are known and can be used without modifications for optimization on quotient versions of these manifolds.

We thus have all optimization primitives for total manifolds, quotient manifolds, and equivalence between quotient manifolds and manifolds from the main text, which allows us to perform optimization on $\varrho_{n, r}$, $C_{n, r}$ and ${\rm POVM}_{n, r}$.

\section{Complexity of algorithms and comparison with other libraries}
\label{app:complexity}

In this appendix, we discuss questions of scalability of optimization algorithms presented in the QGOpt library and compares the QGOpt library with other frameworks. To address the scalability of optimization algorithms, one needs to estimate the asymptotic complexity of primitives used in those algorithms. Table \ref{table:complexity} shows the complexity of optimization primitives for all manifolds.

Let us compare the complexity of algorithms from the QGOpt library with some state of the art algorithms in quantum technologies. For example, let us consider quantum channel tomography, which can be implemented via optimization on $C_{n, r}$. Under the assumption that a particular algorithm uses all the optimization primitives, one step of an optimization algorithm scales like $O(n^3r)$, where $n$ is the dimension of a Hilbert space and $r$ is Kraus rank (see Table~\ref{table:complexity}). In general, the Kraus rank $r$ is equal to $n^2$, which means that the maximal complexity is $O(n^5)$; however, if we have prior information that $r$ is small, then one can significantly reduce the complexity of an algorithm. One can compare the one-step complexity of Riemannian-optimization-based algorithms for quantum channel tomography with the one-step complexity of an algorithm suggested in \cite{knee2018quantum} that is based on the orthogonal projection on the set of CPTP maps. In turn, the orthogonal projection on the set of CPTP maps is implemented through repeated averaged projections on CP and TP sets of maps. The projection on the CP set has complexity $O(n^6)$ that is larger than the complexity of Riemannian-optimization-based algorithms.

\begin{table}[h]
    \centering
    \footnotesize
    \begin{tabular}{| m{1.5cm} | m{2cm}  m{2.5cm}  m{2cm}  m{2cm}  m{2.5cm} |}
     \hline
     Manifold & Retraction & Vector transport & Riemannian gradient & Inner product & Projection \\
     \hline
        ~& ~& ~& ~& ~& ~\\
     $C_{n, r}$& $O(n^3r)$ & $O(\max(n^2r^2,n^3r))$ & $O(n^3r)$ & $O(n^2r)$ & $O(\max(n^2r^2,n^3r))$\\
        ~& ~& ~& ~& ~& ~\\
     ${\rm St}_{n,p}$ & For QR and SVD retractions $O(np^2)$, for Cayley retraction $O(n^3)$ & $ O(np^2) $ & $O(np^2)$ & For Euclidean metric $O(np)$, for canonical metric $O(np^2)$ & $O(np^2)$\\
        ~& ~& ~& ~& ~& ~\\
    ${\rm POVM}_{m,n}$ & $O(mn^3)$ & $O(mn^3)$ & $O(mn^3)$ & $O(mn^2)$ & $O(mn^3)$\\
        ~& ~& ~& ~& ~& ~\\
    $\varrho_{n, r}$ & $O(nr)$ & $O(nr^2)$ & $O(nr)$ & $O(nr)$ & $O(nr^2)$ \\
        ~& ~& ~& ~& ~& ~\\
    $H_n$ & $O(n^2)$ & $O(n^2)$ & $O(n^2)$ & $O(n^2)$ & $O(n^2)$ \\
        ~& ~& ~& ~& ~& ~\\
    $\mathbb{S}_{++}^n$ & $O(n^3)$ & $O(n^3)$ & $O(n^3)$ & $O(n^3)$ & $O(n^2)$\\
        ~& ~& ~& ~& ~& ~\\
    \hline
    \end{tabular}
\caption{Complexity of optimization primitives for all manifolds implemented in the QGOpt library.}
\label{table:complexity}
\end{table}

Let us also compare the QGOpt library with other libraries for Riemannian optimization. Table \ref{table:libraries} shows a list of some related libraries. One can see that QGOpt suits best quantum technologies problems in terms of the number of quantum manifolds.

\begin{table}[h]
    \centering
    \footnotesize
    \begin{tabular}{| m{2cm} | m{3.75cm}  m{8cm} |}
     \hline
     Library & Language & Specific ``quantum'' manifolds \\
     \hline
        ~& ~& ~\\
     QGOpt & Python & Manifolds of density matrices, POVMs, Choi matrices, complex Stiefel manifold\\
        ~& ~& ~\\
     Manopt & Matlab, Python, Julia & POVMs, complex Stiefel manifold\\
     ~& ~& ~\\
     Geoopt & Python & None\\
     ~& ~& ~\\
     mctorch & Python & None\\
        ~& ~& ~\\
     \hline
    \end{tabular}
\caption{Comparison of the QGOpt library with other libraries for Riemannian optimization in terms of the supported ``quantum'' manifolds.}
\label{table:libraries}
\end{table}


\end{appendix}




\nolinenumbers

\end{document}